\documentclass[12pt, a4paper]{article}

\usepackage{graphicx}
\date{}                                                                                
\begin{document}
                                                                                
\title{L\'{e}vy statistical fluctuations from a Random Amplifying Medium}
\author{Divya Sharma, Hema Ramachandran and N. Kumar\\
\it{Raman Research Institute, Sadashivanagar, Bangalore 560 080,
India.}}
\maketitle
                                                                                
\begin{abstract}
We report the studies of emission from a novel random amplifying 
medium that we term a ``L\'{e}vy Laser'' due to the non-Gaussian
statistical nature of its emission over the ensemble of random 
realizations. It is observed that the amplification
is dominated by certain improbable events that are ``larger than
rare'', which  give the intensity statistics a L\'{e}vy like ``fat tail''.
This, to the best of our knowledge, provides the first
experimental realization of L\'{e}vy flight in optics in a random
amplifying medium. 
\end{abstract}

\section{Introduction}
\indent \hspace{7mm}
Random variables having orders-of-magnitude large values, but with
correspondingly orders-of-magnitude small probabilities for their
occurrence, are known to give non-Gaussian statistics for their
fluctuations - the L\'{e}vy statistics \cite{Paul}. For
these {\it larger-than-rare} events the variance diverges, and a single
large event may typically dominate the sum of large number of such
random events. Many physical examples of
L\'{e}vy statistics, or the L\'{e}vy flights, are realized in 
nature, for example, strange kinetics \cite{Shlesinger}, anomalous diffusion 
in living polymers \cite{Ott}, subrecoil laser 
cooling \cite{BardouPRL}, rotating fluid flow \cite{Swinney} and 
interstellar scintillations \cite{Boldyrev}.

In the present work a Random Amplifying Medium (RAM) is shown 
to provide yet another
example of L\'{e}vy statistics of some physical interest. In a RAM, 
light scattering, which is usually
considered detrimental to laser action, can infact, lead to enhanced
amplification and hence to lasing. Here we report 
some analytical and experimental results on the anomalous fluctuations
of emission from a RAM pumped beyond a threshold of gain.
More specifically, we show that in the classical diffusive regime, 
as obtaining in our systems, there is a crossover from the Gaussian 
to the L\'{e}vy statistics for the emission intensity over the 
ensemble of realizations of the random medium. Also, the 
associated L\'{e}vy exponent decreases with the increasing gain. An 
interesting finding is that the L\'{e}vy-statistical
fluctuations are enhanced by embedding an amplifying fiber into the 
particulate RAM. We also briefly discuss the nature of these 
fluctuations as distinct from those observed in transport
through passive random media.

A RAM normally consists of an active
bulk medium, like an optically pumped laser dye solution (for example
Rhodamine in methanol) in which point-like (particulate) scatterers 
(rutile (TiO$_{2}$) or polystyrene microspheres) are randomly 
suspended [7-17].
Unlike the case of a conventional laser with external cavity mirrors 
providing resonant feedback, in a RAM it is the multiple scattering 
of light that provides a non-resonant distributed feedback (Fig 1), and hence 
the mirrorless lasing. The enhanced path-lengths within the 
random medium may arise due to classical diffusion resulting from 
incoherent scattering (dilute suspension of scatterers in a dye) [8-12,16], 
or due to incipient wave localization with strong coherent 
scattering (for example semiconductor powder ZnO, GaN) [13-15]. 
In general, in a RAM operating in the incoherent diffusive regime, 
greater the refractive-index mismatch, greater is the 
diffusive path-length enhancement, and hence greater the amplification. 
A clear signature of lasing in a RAM is the drastic spectral narrowing 
(gain narrowing) of the emission from the system above a well defined 
threshold of pump power. In the dye-scatterer system, the threshold 
of the pump power, at which the emission linewidth collapses from a 
few tens of nanometer to a few nanometers, is almost two orders of 
magnitude smaller in the system with scatterers than the one 
without. Further, with the increase in the scatterer concentration, both, 
the linewidth and the pumping threshold are observed to decrease 
drastically. The selection of the lasing wavelength, however, arises 
here as a result of an optimization involving, for example, the 
wavelength-dependent diffusion coefficient (or the localization 
length scale) and the spectral profile of the pumped dye. 

An important aspect of the random lasing is that for a high gain 
(pumping) the randomness of the amplifying medium makes the emission 
fluctuate strongly over the different microscopic realizations 
(complexions) of the randomness of the medium. This shows up as non-self 
averaging fluctuations of the observed lasing intensity as the medium 
is varied over its random realizations, for example, by tapping the 
cuvette containing the RAM. 
(This, of course, is quite different from the inherent photon 
statistics of fluctuations in time observed for a given complexion 
\cite{Zacharakis}). Normally, i.e., for passive random media, 
these ``sample-to-sample'' fluctuations are Gaussian in nature. 
In this work we will be concerned with fluctuations in a RAM only.

To fix the idea, consider a RAM with the scatterers dispersed densely
and randomly in an amplifying continuum. A spontaneously emitted photon
is expected to diffuse with a diffusion constant $D = (1/3) c\ell$,
where $\ell$ is the elastic mean free path and $c$ is the speed of light in the
medium. We assume classical diffusion as $\ell/\lambda \gg 1$ in our 
case, where, $\lambda$ is the optical wavelength. As the photon diffuses 
and eventually escapes, it undergoes
amplification, or gain (multiplication) due to the optical pumping, and
the associated stimulated emission, which, of course, does not affect $D$.
Assuming for simplicity, a spherical RAM (radius `$a$'), 
illuminated uniformly by a short pump-pulse at time $t = 0$, the
probability of escape of a photon from the surface at $r = a$, per unit time
at time $t$ is given by (the first-passage probability density)
\begin{equation}
p_I(t) = - \frac{\partial}{\partial t} \int_0^a \rho(r,t) 4\pi r^2 dr ~,
\end{equation}
where, $\rho(r,t)$ is the probability density of the diffusing photon,
emitted spontaneously at time $t=0$ anywhere within the sample with a uniform
initial probability density ($\rho_0$). Simple solution for the diffusion
problem (with the absorbing boundary condition at $r = a$) gives
\begin{equation}
\rho(r,t) = \rho_0 \sum_{m=1}^{\infty} \big(\frac{2a}{\pi m}\big ) (-1)^{m+1}
\cdot \frac{sin (\pi m r/a)}{r} e^{-\frac{\pi^2m^2}{a^2}Dt} 
\end{equation}
giving straightforwardly
\begin{equation}
p_I(t) = \rho_0 \sum_{m=1}^{\infty} 8aD e^{-\frac{\pi^2m^2}{a^2}Dt}
\end{equation}
Now, the arc path-length traversed in the diffusion time $t$
is $ct$ giving a gain factor $g = e^{ct/\ell_g}$, where $\ell_g$ is
the gain length for the RAM. This at once gives, with change of 
variable, the probability distribution for the gain $p_g(g)$ as
\begin{equation}
p_g(g) = \sum_{m=1}^{\infty} \big(\frac{\rho_0 8 a D \ell_g}{c}\big )
\frac{1}{g^{1+\alpha_m}}  \equiv \sum_{m=1}^{\infty} 
\big(\frac{8 \rho_0}{3}\big )
(a\ell \ell_g) \frac{1}{g^{1+\alpha_m}}
\end{equation}
with $\alpha_m = m^2 \big(\frac{\pi^2\ell\ell_g}{a^2}\big )$
$\equiv $ the m$^{th}$ L\'{e}vy exponent.
Thus, with increasing pumping (decreasing gain length $\ell_g$), the
exponent $\alpha_m$ decreases, the tail becomes fatter, and the 
variance of $g$ diverges for $\alpha_m < 2$, that happens first 
for $m=1$, i.e., for $(\frac{\pi^2\ell \ell_g}{a^2}) < 2$.
This leads to the crossover from a finite variance (Gaussian) to a  
divergent variance (L\'{e}vy) limit. This essentially describes the onset 
of L\'{e}vy fluctuations as we increase optical pumping. It is 
idealized in that only the photons spontaneously emitted at time
$t=0$ are considered. These are amplified most anyway, and dominate
the intensity at time $t$ observed, for large gains (high pump powers).
Further, in our granular random media with grain size $\gg \lambda$, 
the random scattering is best described as random refractions at the
interfaces. This can give rise to random closed loops that can trap 
and enhance light as in a resonance. Also, inasmuch as the escape 
rate is linked to the diffusion constant, one can expect the classical
Ruelle-Pollicott resonances giving pronounced structure to the 
fluctuation statistics. We have not addressed these issues here.  
                                                                                      
Before we proceed further (with experiments), let us clarify the
meaning of {\it ``fluctuations''} once more in our context. 
These are statistical
fluctuations over the ensemble of realizations of the randomness (i.e., 
macroscopically identical RAMs).
Of course, we can invoke physically the idea of ergodicity and identify
these fluctuations as unfolding in different parametric contexts. Statistical
fluctuations of transmission/conductance though passive random media are,
of course, well known \cite{Kumar}, where, for a macroscopic sample,
the classical fluctuations are small relative to the wave-mechanical 
(or quantum)
fluctuations due to coherent interference effects. In the present case of
strictly classical diffusion ($\ell \gg \lambda$), the anomalously 
large fluctuations are due entirely to the amplification inherent to a RAM.

The system that we have experimentally studied is a novel RAM, which
we term the F-RAM (Fiber-Random Amplifying Medium), inasmuch 
as the active medium is a random aggregation of segments of dye-doped 
amplifying (one-dimensional) fibers (Bicron, red 
fluorescent optical fiber) in a passive  
medium of air, granular starch etc. (Fig 2). These plastic 
fibers fluoresce in the 
orange-red when pumped by green light that enters the fibers through 
their cylindrical surfaces anywhere along their lengths. The emitted 
fluorescent light is mainly guided along the length, and it emerges 
from either end amplified by a factor that increases exponentially 
with the length 
of travel through the fiber. While the random aggregation of the
amplifying fibers itself provides some scattering, the latter
was enhanced in our experiments by the addition of passive scatterers  
like non-active fiber pieces or granular starch. Thus, the diffusion 
proceeds via random scattering and wave-guidance.

Our initial experiments studied the emission from an F-RAM,
made of amplifying fibers crushed to sub-millimeter sizes, both with
and without long pieces of amplifying fibers embedded in it. 
Additionally, these were compared with an
F-RAM consisting of long pieces of amplifying fibers embedded in a passive 
scattering medium. These experiments and the observations are described 
in section 2. The Arrhenius cascade model as also the L\'{e}vy 
microscope \cite{BardouArxiv}, to which the observed 
statistics of  
intensity fluctuations bear relevance, is described in section 3. 
The experimental realization of L\'{e}vy lasers, i.e., F-RAMs with tailored
length distribution, exhibiting the sample-to-sample L\'{e}vy intensity 
fluctuations, in the dilute and the dense limits, is described in 
section 4. Section 5 concludes the work. 

\section{Experiments in F-RAM}
\indent \hspace{7mm}
An F-RAM consisting of amplifying fibers crushed to 
sub-millimeter sizes
(which serve both to amplify and scatter the light), 
was contained in a glass cuvette of size 1 cm$\times$1 cm 
$\times$5 cm. This was pumped by 10 ns, 26 mJ pulses at 532 nm
from a frequency doubled Nd:YAG laser (Spectra Physics). Part of the 
pump beam was split off by a beam-splitter to monitor the
pump intensity that was maintained constant. The emission from the
sample was collected transverse to the pump beam and the spectrum 
analyzed on a PC based spectrometer (Ocean Optics). The schematic 
of the experimental set-up is shown in Fig 3. Lasing action 
was seen from this system above a pump threshold of 22 mJ (Fig 4). 
The complexion of the system was altered i.e. the sample was 
agitated, so that different random configurations were obtained, and 
the resulting emission spectra were recorded. In order to obtain good
statistics, this was repeated till the emission spectra for 420 
different complexions of the sample were obtained.
A histogram was then constructed; that is
the probability $P(I)$ \footnote {$P(I)$ is the number of times 
an intensity was recorded normalized to
the total number of spectra.} of obtaining intensity $I$ 
was plotted as a
function of the intensity. The histograms shown 
for $\lambda = 620~nm$ (emission peak) and $\lambda = 590~nm$ (off-peak)
are both observed to be Gaussian (Fig 5(a,b)). The intensity 
as a function of complexion for these wavelengths 
show small fluctuations (Fig 5(c,d)). 

Ten long pieces (length 6 mm) of amplifying 
fibers were then added to the above F-RAM, and in a similar fashion the spectra 
for 420 different complexions of the sample recorded. A typical spectrum
of this F-RAM is shown in Fig 6. Unlike the earlier case, the histogram  at 
$\lambda = 640~nm$ (peak) shows a marked departure 
from the Gaussian in the form of a long fat tail (Fig 7(a)). 
In addition, the intensity as a function of 
complexion showed sudden large fluctuations (Fig 7(c)). In contrast, at 
$\lambda = 590~nm$ (off-peak), the intensity fluctuations 
remained small (Fig 7(d)) and the histogram Gaussian (Fig 7(b)). 
The departure from the normally observed Gaussian statistics and 
the sudden large intensity fluctuations at the peak emission 
wavelength ($640~nm$) can be explained as 
arising from the few long pieces of amplifying fiber, that, 
in some complexions of the sample, provide large 
gain resulting in the fat tail.
This was verified by studying another system that consisted of a passive
scattering bulk medium (white fiber pieces, length $\sim$ 1 mm), in which 
five pieces of amplifying fiber (length 6 mm) were embedded,  
at pump energy of $\sim$ 12 mJ. The presence of 
the pieces of amplifying fiber, though not visually apparent, is evident 
from the intensity statistics of the emitted spectra  
as a long tail in the histogram at $\lambda = 640~nm$
(Fig 8(a)) and corresponding large intensity fluctuations over 
different complexions (Fig 8(c)). 
On the other hand, the histogram and the intensity fluctuations 
 at $\lambda = 590~nm$ (Figs 8(b,d)) 
show Gaussian statistics. It is thus clear that  
a few long pieces of amplifying fiber 
dominate the emission by their large, but rare, amplification 
so much so that the presence of a few long amplifying pieces 
hidden inside a bulk aggregate of small pieces (active or passive) 
can be inferred from the sample-to-sample fluctuations in the emission 
from the system. This feature may be used to probe a 
relatively long piece of amplifying fiber hidden inside a RAM thus L\'{e}vy
microscope \footnote{The term ``L\'{e}vy microscope'' will become 
clearer after section 3}.

\section{The Arrhenius cascade}
\indent \hspace{7mm}
As the above experiments on F-RAMs indicate that a few large events 
dominate the emission statistics, we are led to the related problem 
of the Arrhenius cascade, which we discuss in brief. The Arrhenius 
cascade studies the time of descent of a particle down an incline 
that has a series of potential wells of varying random depths, $U$, 
occurring with probability
${p_{U}(U) = {\frac{1}{U_o}}~exp({\frac{-U}{U_o}})}$
(${U_{o}}$ is the mean depth) (Fig 9(a): dotted). In a 
well of depth $U$, the particle spends a time $t$, with
${\tau = {\frac{t}{t_o}} = {exp({\frac{U}{kT}})}}$ 
(Fig 9(a): solid). Thus, though deep wells are exponentially improbable,
their presence increases the residence time  exponentially.
It can be shown that in the asymptotic limit, the total time of 
descent follows the power law
${p_{\tau}(\tau) \sim \tau^{(-1-\alpha)}}$ where,
${{\alpha} = {\frac{kT}{U_o}}}$.
For high temperature ($T$), or for $\alpha \geq 2$, the particle has
a fast descent and the resulting distribution ${p_{\tau}(\tau)}$ 
is Gaussian. For $0 < \alpha < 2$, corresponding to intermediate 
or low temperatures, the distribution is L\'{e}vy (Fig 9(b)) and 
the Central Limit Theorem is violated.
                                                                                
To exploit the fact that two functions, one exponentially 
increasing and the other exponentially falling, can combine to
give rise to Gaussian or L\'{e}vy statistics depending on the
relative values of the two exponents, we tailored our F-RAM
system, such that the probability distribution of the lengths of the fibers
was ${p_{\ell}(\ell) = {\frac{1}{\ell_o}}~exp({\frac{-\ell}{\ell_o}})}$,
as shown in Fig 10. (Note that this tailored F-RAM is different from 
those described in section 2 where all long amplifying fiber pieces
were of same length). The amplification within an active fiber results 
in an intensity ${I(\ell) = I_{o}~exp({\frac{\ell}{\ell_g}})}$, or gain
${g_\ell = \frac{I(\ell)}{I_o} = exp({\frac{\ell}{\ell_g}})}$. 
Thus, long fibers, though exponentially
rare, provide exponentially high gain \footnote{Note that the parameters 
${\ell_{o}}$ and ${\ell_{g}}$ in the tailored
F-RAM correspond to ${U_{o}}$ and $kT$ respectively
in the Arrhenius cascade}. 

It can be shown that the probability
distribution of the resultant gain acquired by the photon is given as
${p_g(g) \sim g^{(-1-\nu)}}$ where,
${{\nu} = {\frac{\ell_g}{\ell_o}}}$. It is thus expected 
that $0< \nu < 2$
gives L\'{e}vy intensity statistics and $\nu \geq 2$
Gaussian. We demonstrate experimentally, in the next section, the
crossover from Gaussian to L\'{e}vy as $\ell_g$ is reduced.

\section{Experiments with tailored F-RAM (L\'{e}vy Laser)}
\indent \hspace{7mm}
Experiments were conducted on tailored F-RAMs with N pieces 
(N = 350, 800) of amplifying fibers in passive
scattering media provided by suspension of polystyrene microspheres 
in water (BangsLabs, mean diameter = 0.13 $\mu m$, number density =
$9.357 \times 10^{12}/cc$), granular starch or pieces of white optical 
fiber (non-amplifying, length $\sim$ 0.5 mm to 1 mm). In all 
three systems (contained in glass cuvettes of size 30 mm$\times$ 30 mm
$\times$ 60 mm) were studied in which, the lengths of the 
amplifying fibers ranged from 1 mm to 20 mm and followed an exponential
distribution with $\ell_{o}$ = 5 mm.

As described in section 2, spectra (at pump energies $\sim$ 6-9 mJ)
for $\sim$ 360 different complexions of
each of the systems were obtained and analyzed. The intensity fluctuations and
the corresponding histograms are given in Figs 11 to 13.
These are shown for $\lambda = 645~nm$ and  $590~nm$, 
the former corresponding to the peak emission wavelength, where the gain is
maximum ($\ell_g$ is minimum), and the latter to off-peak
wavelength ($\ell_g$ is large). The histograms of all three systems 
show a L\'{e}vy-like 
fat tail at the peak emission wavelength; therefore these tailored 
F-RAMs are termed L\'{e}vy lasers. 
In contrast, at off-peak wavelengths, the histograms show Gaussian statistics
consistent with the larger value of $\nu$. 
The intensity at the peak wavelength ($645~nm$) as a function 
of complexion showed sudden large jumps, typical of L\'{e}vy flights. 
This feature was absent at off-peak wavelengths.

We now distinguish between the ``dilute'' and the ``dense'' limits 
of the L\'{e}vy Laser. The dilute L\'{e}vy Laser contains a few
pieces of amplifying fibers. A photon originating within a
given piece of amplifying fiber gains in intensity as it traverses
the fiber. Upon exiting the fiber, it diffuses through the passive surrounding
medium and exits the sample with a negligible probability of
encountering another amplifying fiber (Fig 14(a)). The intensity 
collected in the experiment is the sum of various such
intensities - the {\it additive gain}. As discussed earlier, it gives
a power-law for the gain i.e., ${p_g(g) \sim g^{-1-\nu}}$.
Of the systems studied, the case
with N = 350 amplifying fibers in polystyrene
scattering medium corresponds to a dilute system.
The tail of the histogram can be fitted to a power law function 
($g^{-1 -\nu}$) with exponent 1 + $\nu$ = 2.69 i.e., $\nu$ = 1.69.
                                                                                
In the dense L\'{e}vy laser, on the other hand, a photon, upon 
exiting an amplifying
fiber, has a high probability of entering another amplifying fiber and
getting further amplified before finally exiting the sample (Fig 14(b)). 
In such a case, the total intensity (or gain) is {\it multiplicative}
rather than {\it additive}
i.e. ${G = \Pi_{i}~ g_{\ell_i} = \Pi_{i}~exp({\frac{\ell_i}
{\ell_g}})}$,
where, the index, i,  runs over all fibers that a given photon traverses
through, from which we get ${p_x(x) \sim g^{-\nu}}$
where, ${x = ln~g_{\ell_i}}$.
Thus, the dense system with multiplicative gain also gives rise to
a L\'{e}vy distribution, but with a tail that falls off slower than
the dilute system.
Cases with N = 800 amplifying fibers in passive scattering media 
are realizations of dense L\'{e}vy lasers. 
The tails of the histograms can be fitted to the power 
law function ($g^{-\nu}$) with exponents $\nu$ = 0.62 and 1.68, 
for systems with passive scattering medium as
non-active white fiber pieces and granular starch respectively.
                                                                                
\section{Conclusions}
\indent \hspace{7mm}
We have demonstrated a new RAM, namely the F-RAM, that is notably 
different from a conventional RAM in several aspects. 
As opposed to the RAM that has a bulk active medium
(dye solution) with suspended passive point-like scatterers,
an F-RAM, has an active medium that is one-dimensional
(pieces of amplifying fiber) and is suspended in the passive bulk
medium. Further, unlike the conventional RAM,
during its traversal through the passive bulk medium in an F-RAM
the photon does not get amplified.
Consequently, a greater refractive index mismatch between the active
(fiber) and the passive (bulk) media, 
which in the case of RAM leads to greater amplification
due to increased path-length, is likely to 
result under some conditions in just the opposite in an F-RAM, 
as it  enhances scattering 
off the active fiber. We term an F-RAM with a tailored distribution of 
fiber lengths, where long amplifying pieces are exponentially rare, a 
``L\'{e}vy Laser'', because the sample-to-sample intensity fluctuations 
exhibit L\'{e}vy statistics. The ``larger than rare'' amplification in 
such systems
makes feasible a ``L\'{e}vy microscope'' that can pick out the presence
of, and study the characteristics of a long piece of amplifying 
fiber embedded in a bulk of smaller (active or passive) pieces.

\newpage

\begin{figure}
\centering
\includegraphics[width=16cm]{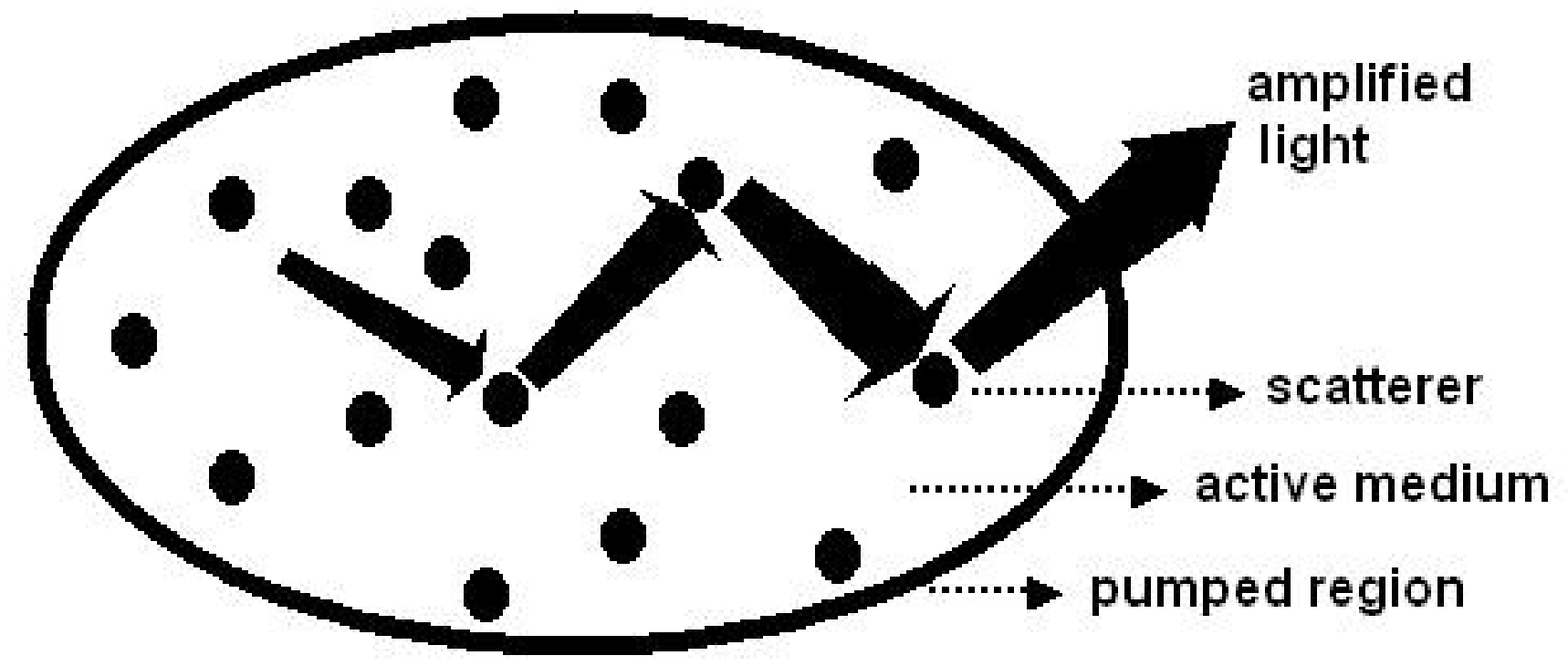}
\caption{\it Schematic of a RAM illustrating amplification due to 
multiple scattering.}
\label{figure1}
\end{figure}

\clearpage

\begin{figure}
\centering
\includegraphics[width=16cm]{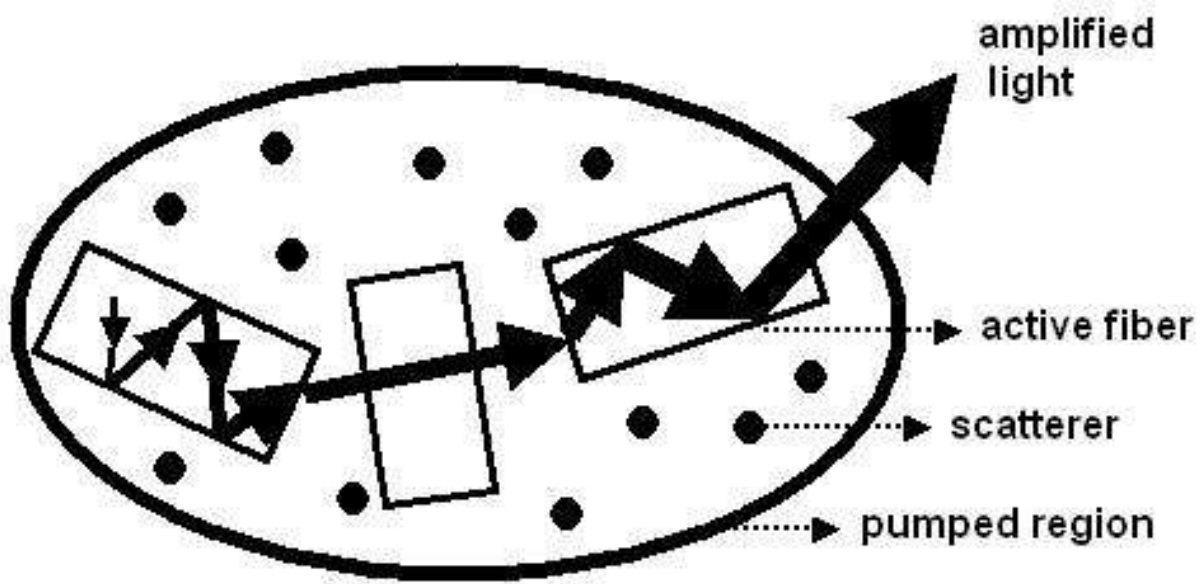}
\caption{\it Schematic of an F-RAM illustrating amplification of light
within the active fibers.}
\label{figure2}
\end{figure}

\clearpage

\begin{figure}
\centering
\includegraphics[width=16cm]{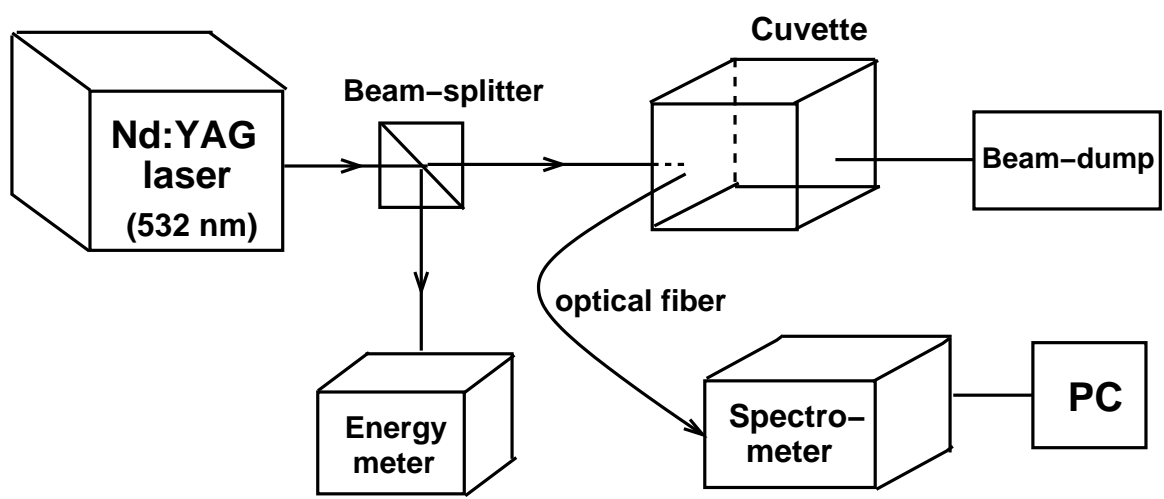}
\caption{\it Schematic of experimental set-up.}
\label{figure3}
\end{figure}
                                                                                            
\begin{figure}
\centering
\includegraphics[width=15cm]{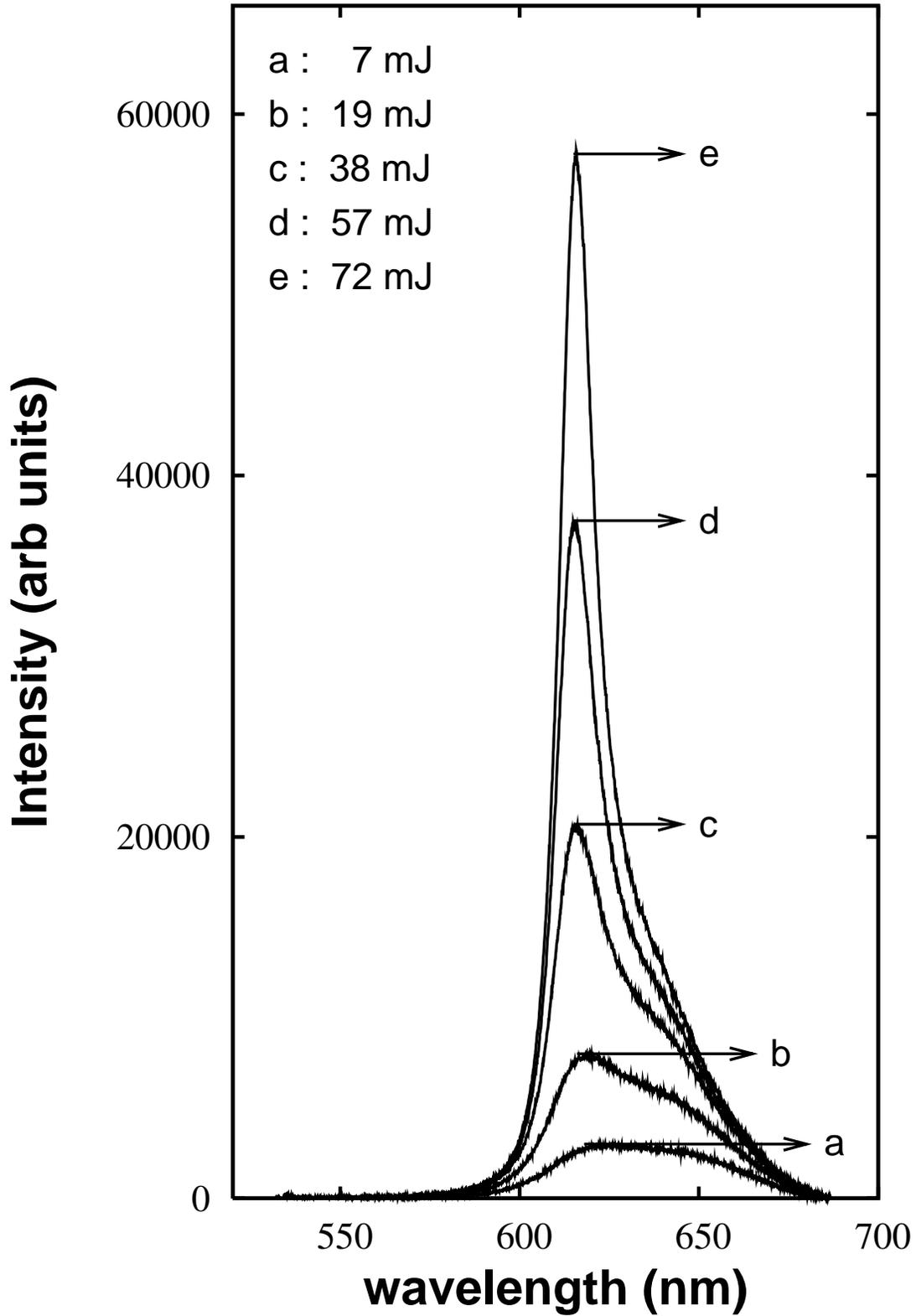}
\caption{\it  Gain narrowing with increasing pump powers in an F-RAM
made of sub-millimeter pieces of active fiber in water.}
\label{figure4}
\end{figure}
                                                                                            
\begin{figure}
\centering
\includegraphics[height=20cm,width=16cm]{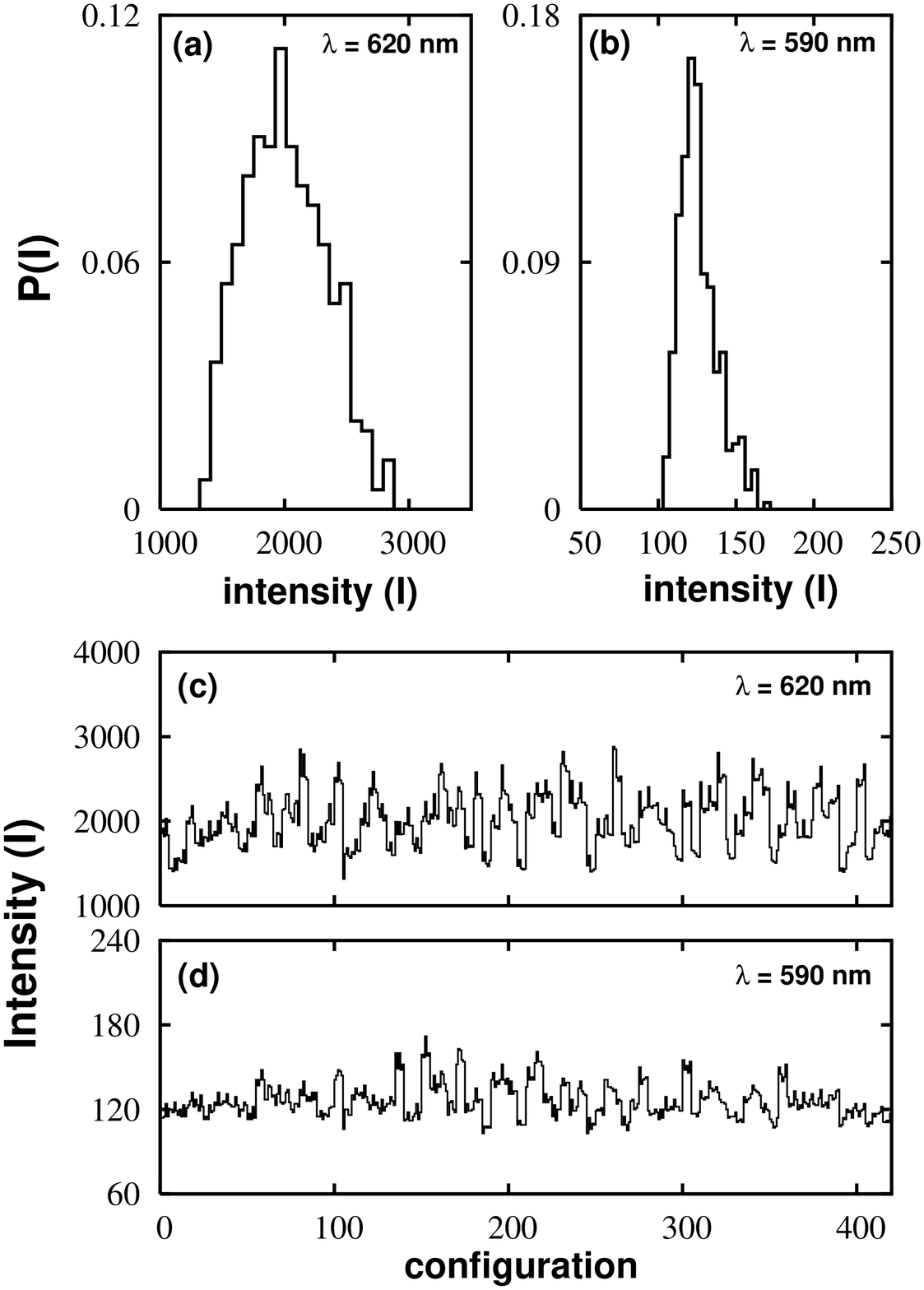}
\caption{\it (a),(b) : Histograms of emission at peak and off-peak wavelengths
respectively for F-RAM consisting of sub-millimeter pieces of active fiber.
(c),(d) : Intensity fluctuations as a function of complexion for the F-RAM at
peak and off-peak emission wavelengths.}
\label{figure5}
\end{figure}
                                                                                            
\begin{figure}
\centering
\includegraphics[height=15cm,width=15cm]{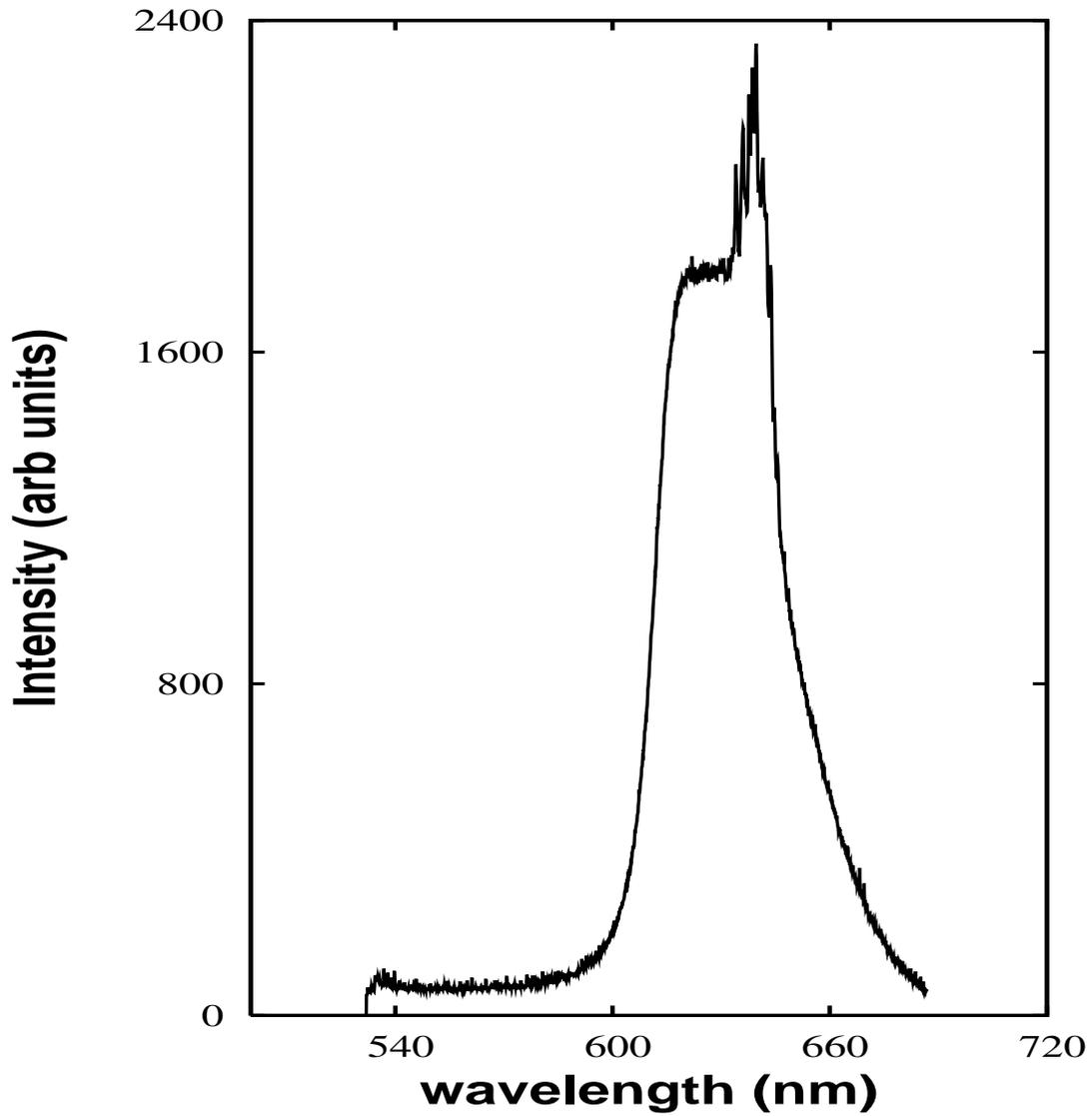}
\caption{\it  Typical spectrum of an F-RAM consisting of ten pieces 
of active fiber of length 6 mm each, embedded in sub-millimeter
pieces of active fiber, at pump energy of $\sim$ 26 mJ.}
\label{figure6}
\end{figure}

\begin{figure}
\centering
\includegraphics[height=20cm,width=16cm]{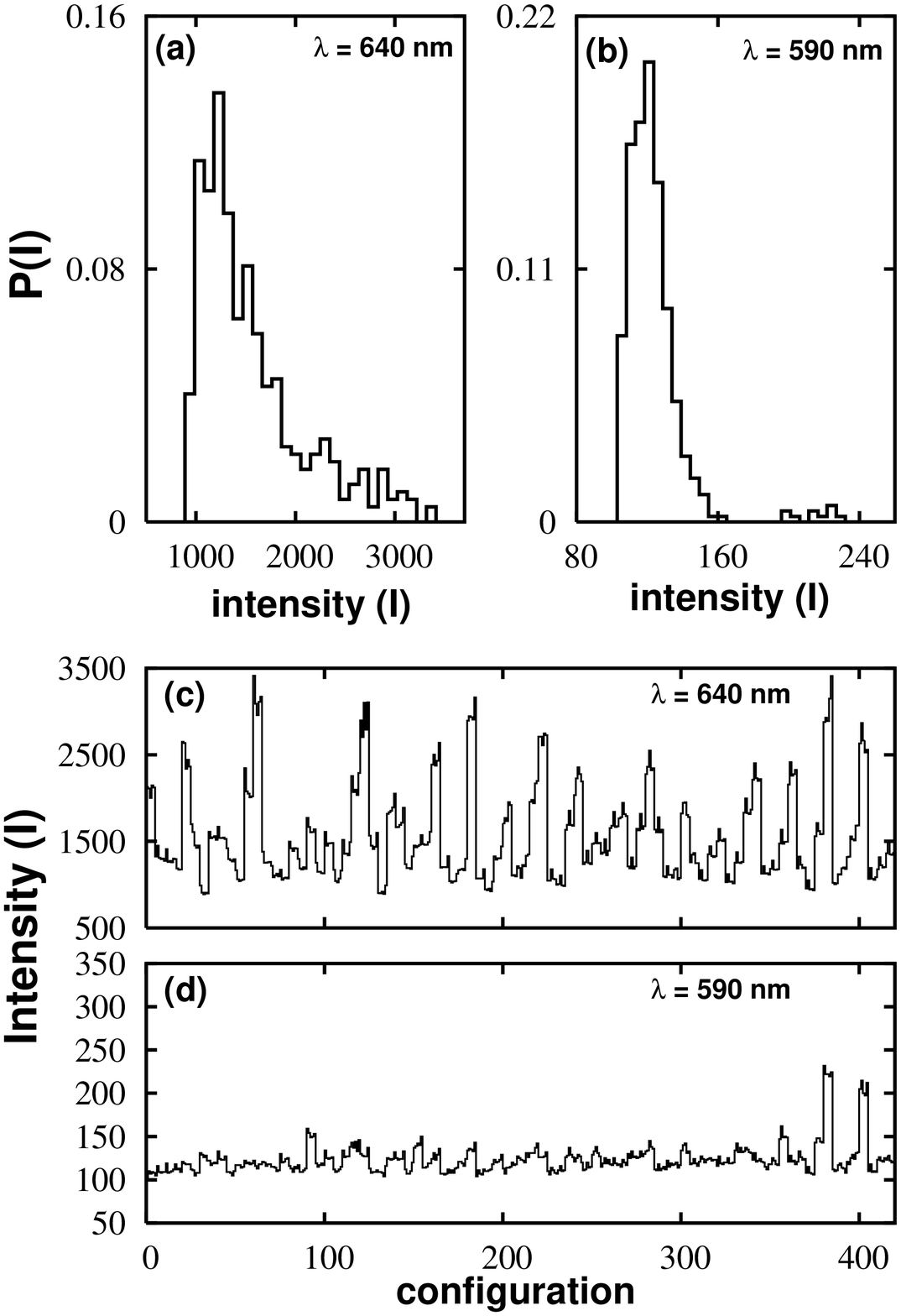}
\caption{\it (a),(b) : Histograms of emission at peak and off-peak wavelengths
respectively for F-RAM consisting of ten pieces of active fiber of length 
6 mm each, embedded in sub-millimeter pieces of active fiber.
(c),(d) : Intensity fluctuations as a function of complexion for the F-RAM at
peak and off-peak emission wavelengths.}
\label{figure7}
\end{figure}
                                                                                                        
\begin{figure}
\centering
\includegraphics[height=20cm,width=16cm]{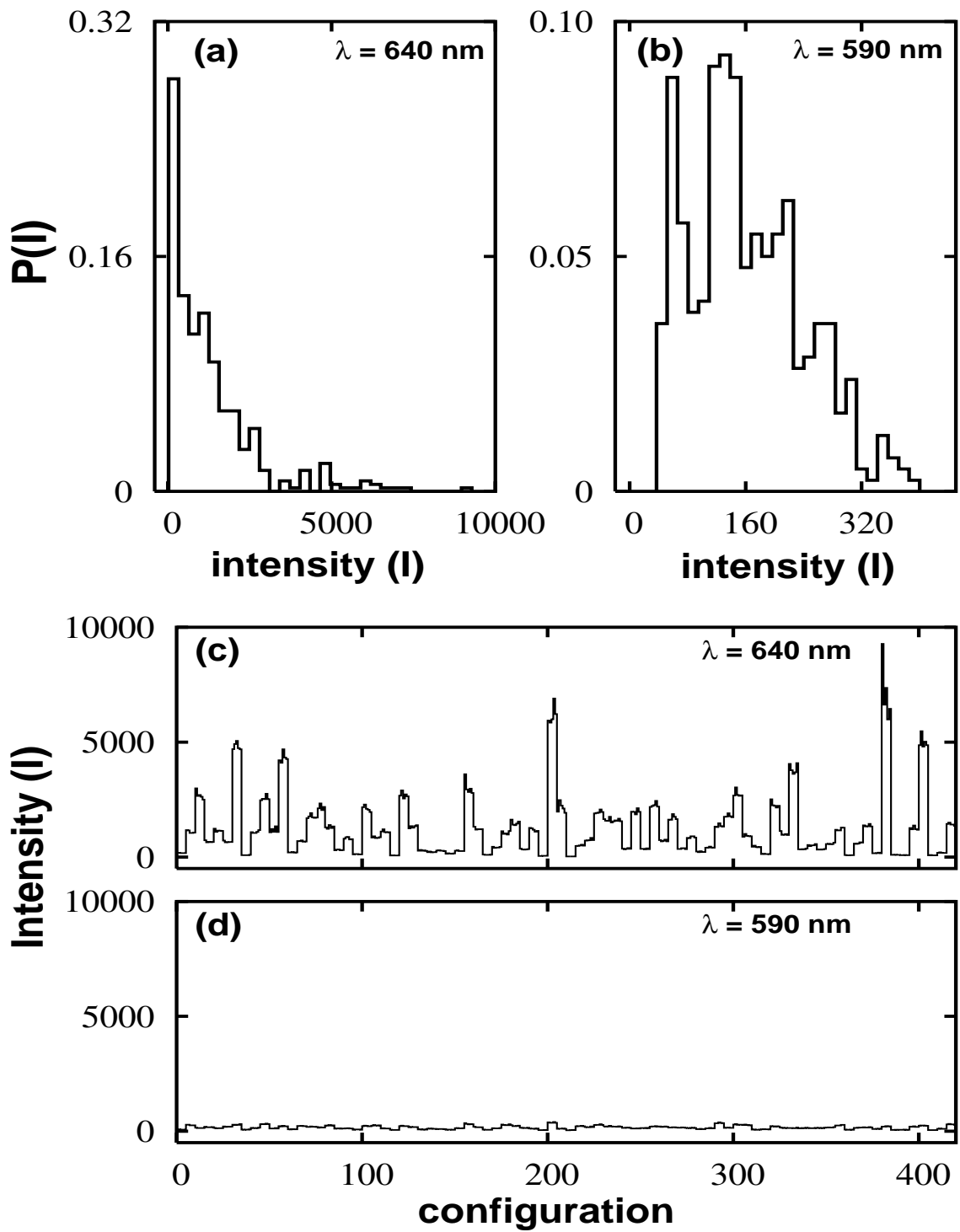}
\caption{\it (a),(b) : Histograms of emission at peak and off-peak wavelengths
respectively for F-RAM consisting of ten pieces of active fiber of length
6 mm each, in a passive scattering medium made of pieces of non-active white fiber of length
$\sim$ 1 mm each. (c),(d) : Intensity fluctuations as a function of complexion for the F-RAM at
peak and off-peak emission wavelengths.}
\label{figure8}
\end{figure}                                                                                                                                                                                                           
\begin{figure}
\centering
\includegraphics[height=8cm]{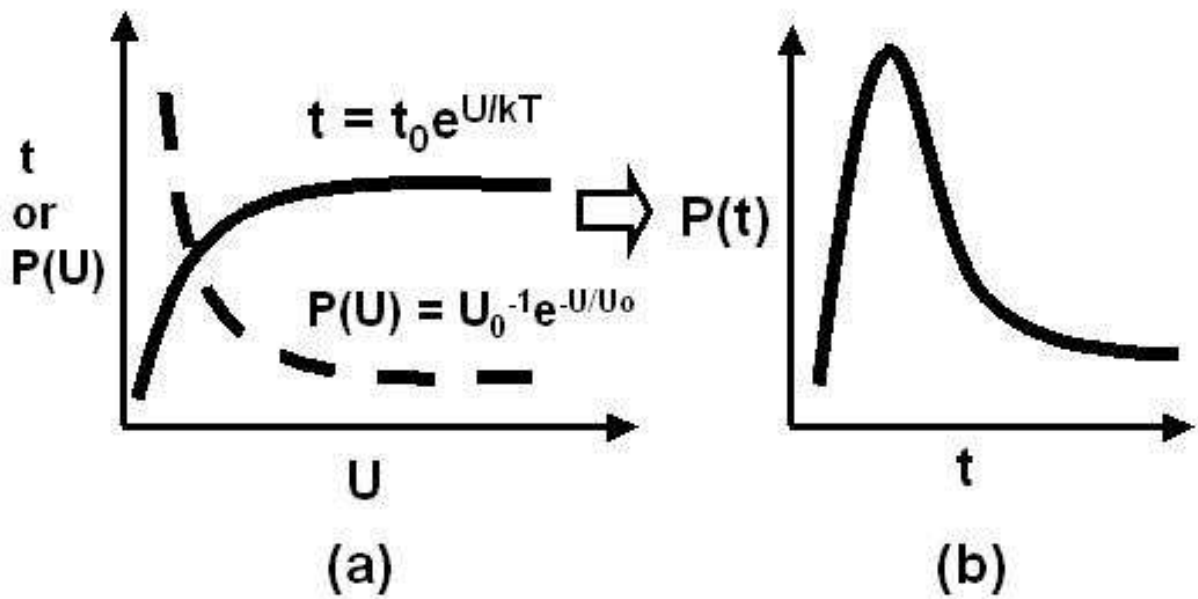}
\caption{\it  (a) : Distribution of depths of potential wells in 
an Arrhenius cascade (dotted) and the time of residence in a potential 
well as function of depth $U$ (solid), (b) : Resulting probability 
distribution of total time of descent.}
\label{figure9}
\end{figure}

\begin{figure}
\centering
\includegraphics[height=15cm]{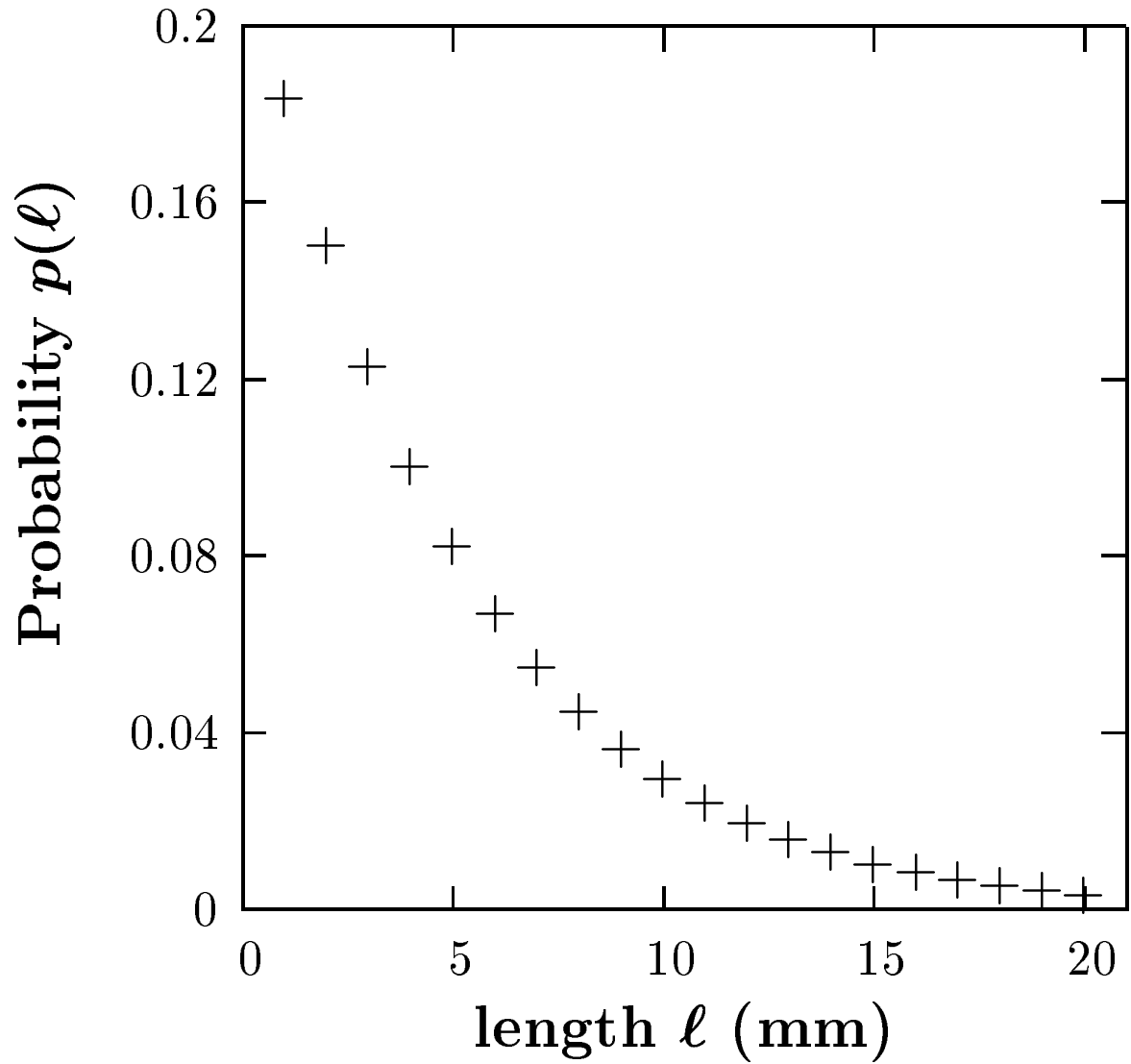}
\caption{\it  Probability distribution function for lengths of pieces
of active fiber in tailored F-RAM : $\ell_o = 5$~mm.}
\label{figure10}
\end{figure}

\begin{figure}
\centering
\includegraphics[height=20cm,width=16cm]{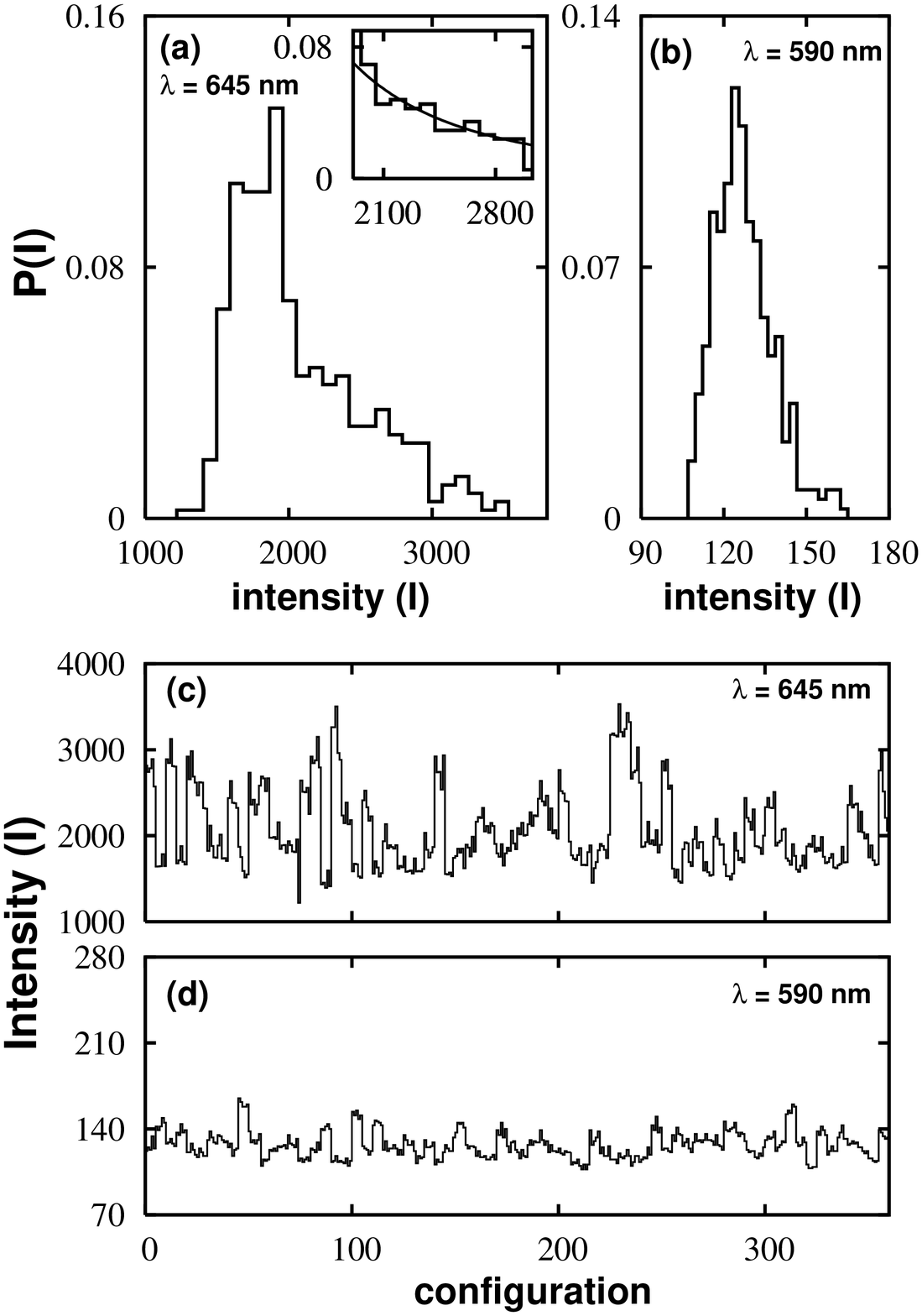}
\caption{\it (a),(b) : Histograms of emission at peak and off-peak wavelengths
respectively for F-RAM consisting of 350 pieces of active fiber (following 
exponential distribution for lengths) in a passive scattering medium provided by
polystyrene scatterers. (c),(d) : Intensity 
fluctuations as a function of complexion for the F-RAM at
peak and off-peak emission wavelengths.}
\label{figure11}
\end{figure}

\begin{figure}
\centering
\includegraphics[height=20cm,width=16cm]{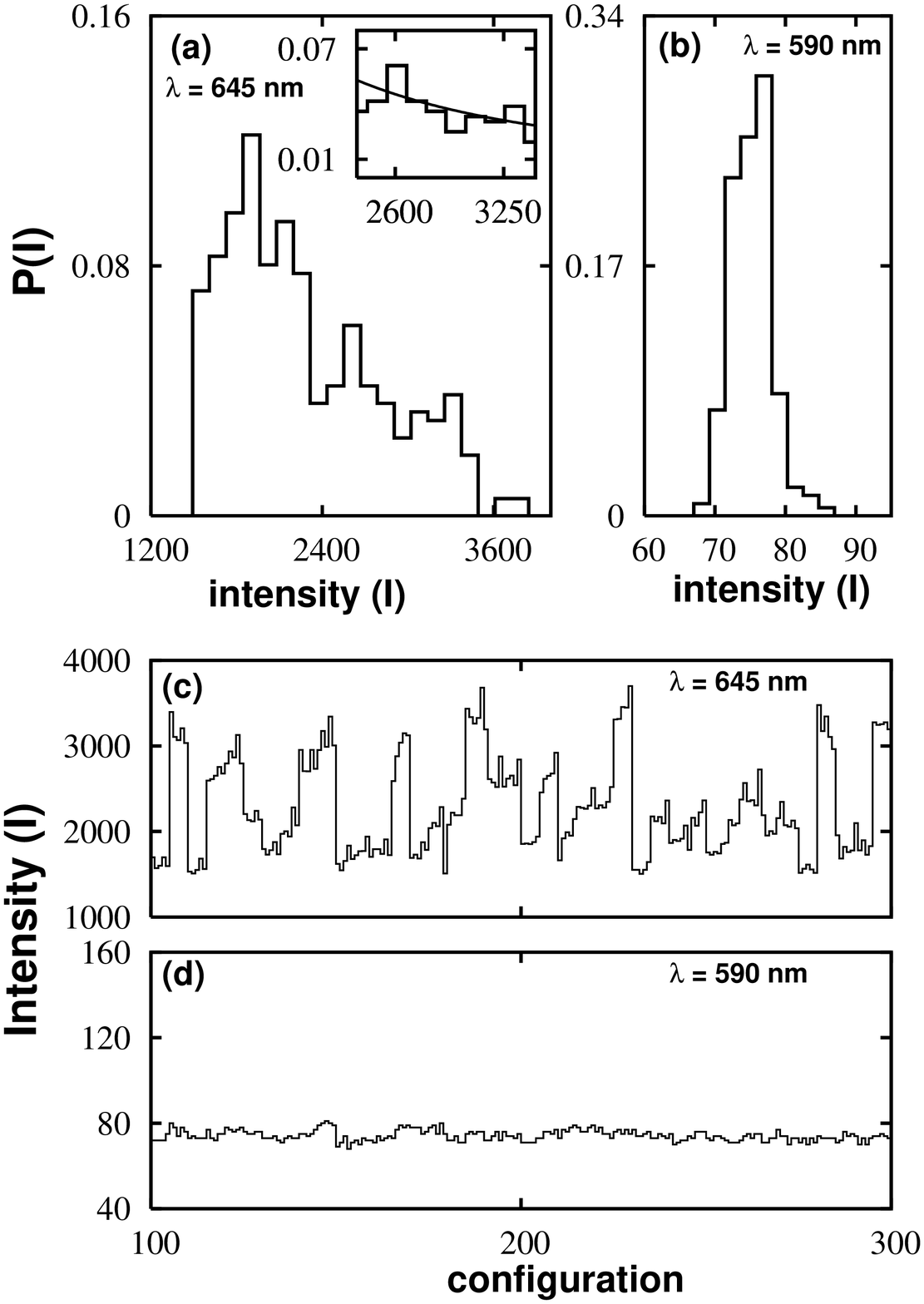}
\caption{\it (a),(b) : Histograms of emission at peak and off-peak wavelengths
respectively for F-RAM consisting of 800 pieces of active fiber (following
exponential distribution for lengths) in a passive scattering medium provided by
granular starch. (c),(d) : Intensity
fluctuations as a function of complexion for the F-RAM at
peak and off-peak emission wavelengths.}
\label{figure12}
\end{figure}

\begin{figure}
\centering
\includegraphics[height=20cm,width=16cm]{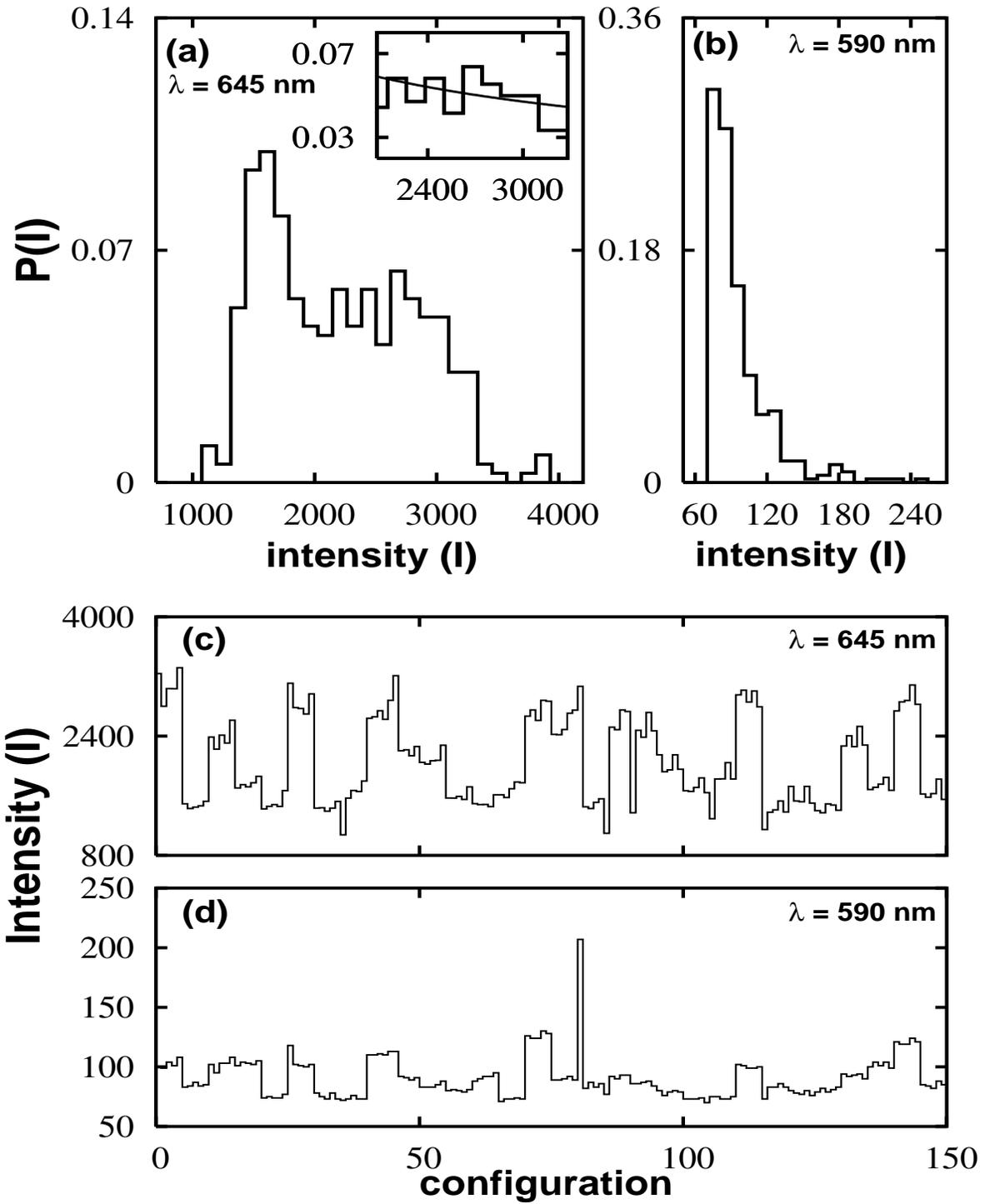}
\caption{\it (a),(b) : Histograms of emission at peak and off-peak wavelengths
respectively for F-RAM consisting of 800 pieces of active fiber (following
exponential distribution for lengths) in a passive scattering medium provided by
non-active white fiber pieces (length $\sim$ 1 mm). (c),(d) : Intensity
fluctuations as a function of complexion for the F-RAM at
peak and off-peak emission wavelengths.}
\label{figure13}
\end{figure}

\begin{figure}
\centering
\includegraphics[height=19cm,width=15cm]{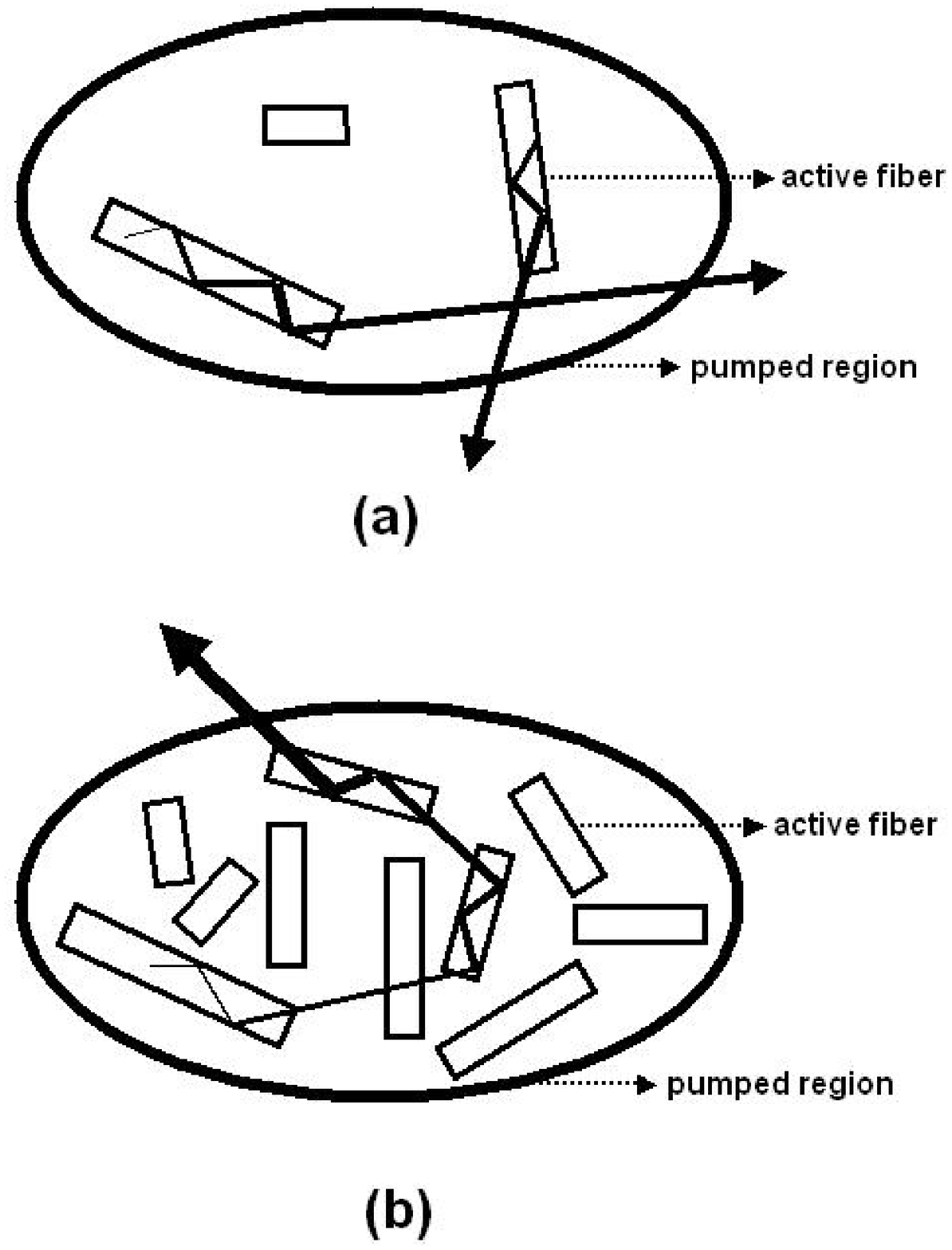}
\caption{\it Schematic of (a) Dilute F-RAM (b) Dense F-RAM}
\label{figure15}
\end{figure}

\end{document}